\documentclass[runningheads]{llncs}
\usepackage{graphicx}
\usepackage{todonotes} 

 \usepackage{comment}
\usepackage{array}
\usepackage{framed}

 \usepackage[colorlinks=true,
        linkcolor=black,
        citecolor=black,
        filecolor=black,
        urlcolor=black,
        bookmarks=true,
        bookmarksopen=true,
        bookmarksopenlevel=3,
        plainpages=false,
        pdfpagelabels=true]{hyperref}

\usepackage{ifthen}

\newboolean{auxinfo}
\setboolean{auxinfo}{true}   

\newboolean{showoutline}
\setboolean{showoutline}{true}

\newboolean{showcomment}

\setboolean{showcomment}{true} 

\usepackage[subtle]{savetrees}

\ifthenelse{\boolean{auxinfo}}
  {}
  {
    \setboolean{showoutline}{false}
    \setboolean{showcomment}{false}
  }

\usepackage{xcolor}
\usepackage{amssymb}

\ifthenelse{\boolean{showcomment}}
   {\newcommand{\nb}[2]{\fcolorbox{gray}{yellow}{\bfseries\sffamily\scriptsize#1}{\sf\small$\blacktriangleright${\em #2}$\blacktriangleleft$}}
   \newcommand{\working}[1]{\fcolorbox{gray}{yellow}{{\bf #1}\emph{\scriptsize---in progress---}}}
   \newcommand{\TBD}[1]{\fcolorbox{gray}{yellow}{{\bf #1}\textbf{TBD}}} 
  }
  {\newcommand{\nb}[2]{}{}
   \newcommand{\working}[1]{}
   \newcommand{\TBD}[1]{} 
  }

\usepackage{todonotes}
\ifthenelse{\boolean{showoutline}}{
	\newcommand{\outline}[3]{
		~\newline 
		\fcolorbox{red}{white}{
			\parbox{\columnwidth}{
				\ifthenelse{\equal{#1}{}}{
					\ifthenelse{\equal{#2}{}}{
						\noindent\colorbox[rgb]{0.65,0.16,0}{\textcolor[rgb]{1,1,1}{\textbf{Outline}}}
					}{
						\colorbox[rgb]{0.65,0.16,0}{\textcolor[rgb]{1,1,1}{\textbf{Outline -- Responsible: #2}}}
					}
				}{
					\ifthenelse{\equal{#2}{}}{
						\noindent\colorbox[rgb]{0.65,0.16,0}{\textcolor[rgb]{1,1,1}{\textbf{#1 page(s)}}}
					}{
						\colorbox[rgb]{0.65,0.16,0}{\textcolor[rgb]{1,1,1}{\textbf{#1 page(s) -- Responsible: #2}}}
					}
				}
				#3
			}
		}
	}
}{
	\newcommand{\outline}[3]{}
}

\newcommand\defauxcomm[1]{
       \expandafter\newcommand\csname #1\endcsname[1]{\nb{#1}{##1}}
       \expandafter\newcommand\csname WK#1\endcsname{\working{#1}}
       \expandafter\newcommand\csname TBD#1\endcsname{\nb{#1}}
    } 
   
\defauxcomm{KS}
\defauxcomm{MN}

\usepackage[normalem]{ulem}
\ifthenelse{\boolean{showcomment}}
    {\newcommand{\strike}[1]{\textcolor{red}{\sout{#1}}}}
    {\newcommand{\strike}[1]{}}
\begin{document}
\title{Teaching Agile Requirements Engineering:
A Stakeholder Simulation with Generative AI}
\titlerunning{A Stakeholder Simulation with Generative AI}
%
\author{Eva-Maria Schön\inst{1}\orcidID{0000-0002-0410-9308}\and
Michael Neumann\inst{2}\orcidID{0000-0002-4220-9641}\and
Tiago Silva da Silva\inst{3}\orcidID{0000-0001-8459-7833} }

\authorrunning{Schön et al.}
%
\institute{University of Applied Sciences Emden/Leer,\\  Emden/Leer, Germany
\email{eva-maria.schoen@hs-emden-leer.de}\\
\and
University of Applied Sciences and Arts Hannover,\\ Hannover, Germany 
\email{michael.neumann@hs-hannover.de}\\
\and
Federal University of S\~ao Paulo, S\~ao Paulo, Brazil\\
\email{silva.tiago@unifesp.br}
}
%
\maketitle              
\begin{abstract}
\textit{Context:} The active involvement of users and customers in agile software development remains a persistent challenge in practice. For this reason, it is important that students in higher education become familiar with good practices in Agile Requirements Engineering during their studies.
\textit{Objective:} Our objective is to enable students to learn how to interact with Generative Artificial Intelligence (GenAI) through the use of a stakeholder simulation with AI Personas, while also developing an understanding of the limitations of AI tools in practical contexts. 
\textit{Method:} In our courses, we employ a stakeholder simulation using GenAI, in which students conduct interviews with AI Personas through a provided meta-prompt. Based on the outcomes of these interviews, students apply agile practices (e.g., story mapping or impact mapping) to document requirements. The use of GenAI is subsequently reflected upon in a structured group discussion.
\textit{Results:} Through this approach, students gain practical experience by applying state-of-the art agile practices for requirements elicitation and documentation while simultaneously developing an understanding of the technical and ethical limitations associated with the use of generative AI. 
\textit{Conclusion:} We have applied this approach over several terms and found that using a meta-prompt provides flexibility, allowing us to remain independent of specific large language model providers.

\end{abstract}
%
%
\section{Introduction}
The disruptive changes triggered by the public availability of Generative Artificial Intelligence (GenAI) at the end of 2022, following the release of ChatGPT~\cite{OpenAI.2022}, have not yet reached their peak. In recent years, experimentation with this technology has strongly shaped higher education, with lecturers and students jointly gaining hands-on experience and learning which applications are effective and where limitations emerge. GenAI has fundamentally altered how teaching and learning take place, with particular implications for higher education in the field of software engineering~\cite{Neumann2023}. One insight has become increasingly clear: without sufficient knowledge of AI literacy and ethical considerations, this powerful technology is often misused, since existing guidelines are often ignored~\cite{Bischof.2025}.

Motivated by this observation, our goal is to integrate GenAI into the classroom in a meaningful and pedagogically grounded manner. To this end, we have been using GenAI for teaching Agile Requirements Engineering (RE) for several terms (since winter term 24/25, Nov. 2024). Agile RE follows an iteratively approach across the entire product development lifecycle~\cite{Sporsem.2025}. Requirements are treated as hypotheses that are continuously refined through ongoing feedback from stakeholders and users~\cite{Schon2017f}. Despite this iterative nature, practitioners encounter various challenges when performing RE activities in real-world settings~\cite{Fernandez2017}. Recent research further indicates that the active involvement of users and customers remains a persistent challenge in practice~\cite{Kollmorgen.2025}. However, we use stakeholder simulations based on \textit{AI Personas} to demonstrate the iterative nature of Agile RE and to address the advantages and disadvantages of using GenAI in this context.

The persona method has long been established in Agile RE~\cite{Schon2017f} and Human-Centered Design \cite{iso9241} to address the needs of different target groups during the development of interactive systems. \textit{AI Personas} (also known as "synthetic users"~\cite{Pehar2025}) extend this approach by implementing personas through GenAI chatbots such as ChatGPT. These AI Personas are synthetic constructs designed to emulate human-like characteristics and behaviors tailored to specific roles within GenAI systems~\cite{Karimova.2025}. However, the use of AI Personas also entails limitations. Since the data underlying these AI Personas originates from the training data of large language models (LLMs), they may reflect inherent biases. Consequently, the use of AI Personas for the actual development of interactive systems remains limited, making their application particularly suitable for educational and reflective contexts.

This paper is structured as follows: First, we provide an overview of our teaching case and offer teaching tips for other lecturers on how to apply it in Section~\ref{sec:TeachingCase}. Section~\ref{sec:PracticalImplications} reflects on our experiences with its use and share our learnings related to didactic design and technical implementation. At the end, we close the paper with a conclusion and outline our future work in Section~\ref{sec:Conclusion}.

\section{Teaching Case and Teaching Tips}
\label{sec:TeachingCase}

\subsection{Example of Embedding the Teaching Case}
\label{sec:EmbeddingTeachingCase}
In this Section, we describe our teaching case on stakeholder simulation using GenAI, which we have already applied in several modules. The following section illustrates a concrete implementation of the case within a specific module. The here presented teaching case was applied during the 2024/2025 winter term at the University of Applied Sciences and Arts Hannover. An overview of the instructional setting and implementation conditions is presented in Table~\ref{tab:teaching_case}.

\begin{table}[h]
\centering
\renewcommand{\arraystretch}{1.3}
\begin{tabular}{|p{3.5cm}|p{8cm}|}
\hline
\textbf{Topic} & \textbf{Description} \\ \hline

Degree programs & 
Bachelor of Science (B.Sc.):
\begin{itemize}
    \item Business Information Systems (BIS)
    \item E-Government (EGOV)
\end{itemize}
\\ \hline

Module & 
Requirements Analysis (3rd term of studies)
\\ \hline

Learning Objectives of Module & 
The students are able to analyze requirements for software products by applying the object oriented paradigm using Unified Modeling Language.
\\ \hline

Students & 
The group size is around 65 students for BIS and 30 students for EGOV. The students have diverse backgrounds in terms of profession and school examination.
\\ \hline

Implementation of Teaching Case & 
The teaching case is implemented as an impulse workshop. Attendance at the workshop is not mandatory.
\\ \hline

Duration of case & 
The workshop is planned for 90 minutes.
\\ \hline

Term of application & 
Winter term 2024/2025
\\ \hline

\end{tabular}
\caption{Overview of the Teaching Case}
\label{tab:teaching_case}
\end{table}

As part of a teaching case implemented in the winter term 2024/2025 at the University of Applied Sciences and Arts Hannover, students were trained to elicit and analyze requirements for a mobile public administration app. The objective was to prepare prospective requirements engineers to work with different user types and to understand the challenges of digital transformation in the public sector. The scenario focused on typical citizen services, such as applying for identity documents or registering a change of residence, while reflecting the scepticism and change-resistance often found in public administrations. At the same time, the involved stakeholders were intrinsically motivated to support the requirements engineer and provide realistic requirements.

Students conducted elicitation sessions with three predefined stakeholders representing different user perspectives: A public administration specialist, an accessibility-focused citizen, and a power user from a large city. Each stakeholder
provided concise requirements for an appointment booking feature, expressed preferences, and occasionally showed resistance by posing counter questions or referencing failed app functions from well-known administrations. The students were expected to respond empathetically, stay within the defined project context, and steer the conversation back to the elicitation goal when needed.

Students applied various elicitation methods (e.g., interviews, focus groups), analyzed the resulting requirements, and documented them in user story format in a product backlog. The exercise enabled them to practice realistic stakeholder
communication, handle conflicting expectations, and translate elicited information into structured agile requirements.

\subsection{Learning Objectives}
\begin{figure}[htbp]
\centering
\includegraphics[scale=0.04]{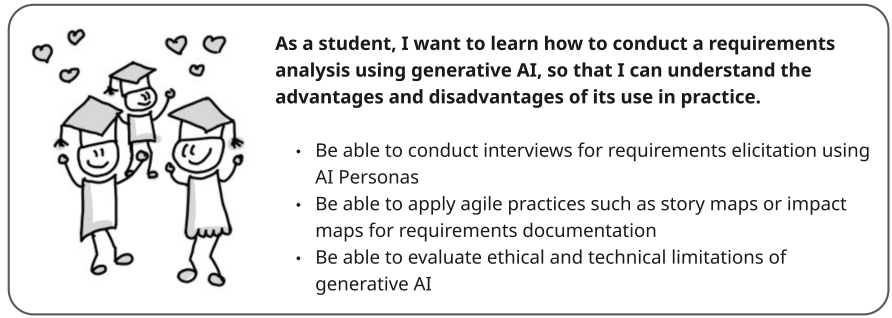}
\caption{Learning objectives formulated in form of a user story}
\label{fig_LearningObjectives}
\end{figure}

In our courses, we often formulate learning objectives as user stories~\cite{Cohn2004} to clearly express the relevance of the tasks from the students’ perspective. The acceptance criteria are described in the form of learning outcomes aligned with the revised learning objectives taxonomy of~\cite{Anderson2001}. Figure~\ref{fig_LearningObjectives} presents the learning objectives for the stakeholder simulation using generative AI. This user story is discussed with the students at the beginning of the exercise to ensure that the learning objectives of the activity are clearly understood. 

\subsection{Teaching Tips for Application of the Teaching Case}

\begin{figure}[htbp]
\centering
\includegraphics[scale=0.073]{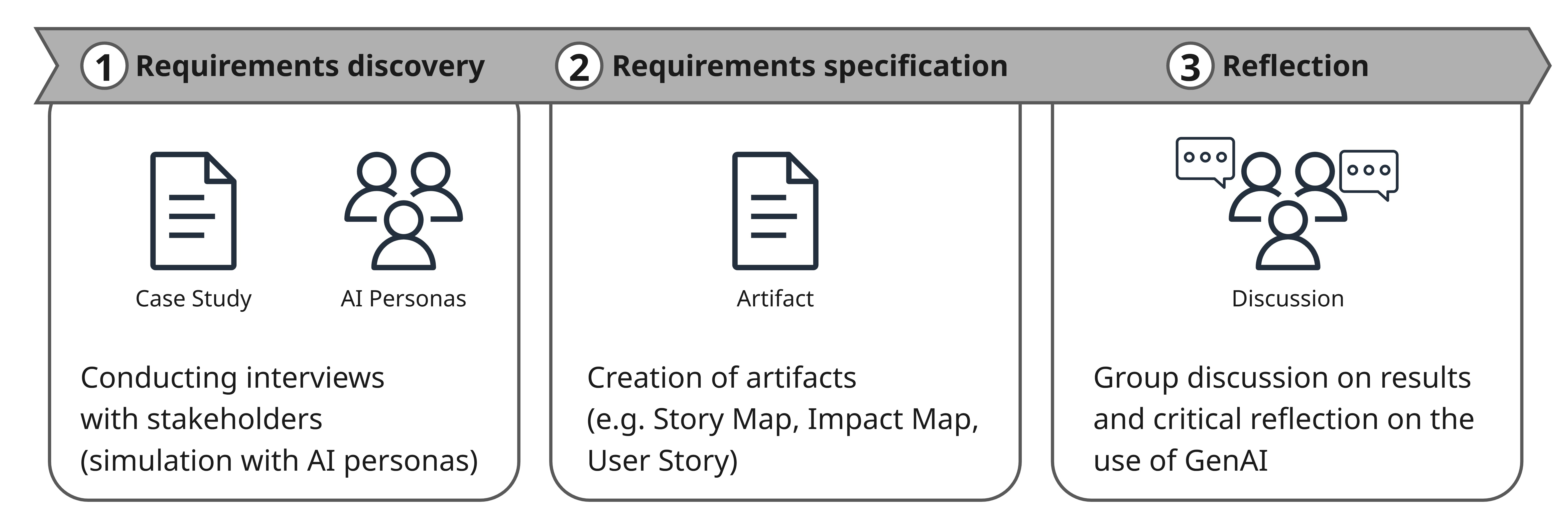}
\caption{Overview of the elements of the teaching case}
\label{fig_TeachingCase}
\end{figure}

This section outlines teaching tips for the application of the teaching case. The teaching case can be applied to both on-campus and online study programs. The stakeholder simulation consists of three elements that are described in the following (see Figure~\ref{fig_TeachingCase}).

\textit{1. Requirements Discovery}. In the first step, students receive the task description. This includes a case study (\textit{e.g.,} the development of a mobile application for the digitalization of citizen-oriented public services). In addition, students are provided with a meta-prompt that can be used with any GenAI chatbot, such as ChatGPT. Once the meta-prompt is entered into the chatbot, it simulates the preconfigured AI Personas representing the relevant stakeholders. Students can then conduct interviews with these stakeholders and ask questions (\textit{e.g., What problems do citizens experience with the current digital service offerings?}). The stakeholders respond according to their predefined roles configured in the meta-prompt.

\textit{2. Requirements Specification}. In the next step, students are asked to transfer the requirements collected during the interviews into appropriate artifacts. Various artifacts from Agile RE can be used for this purpose. Based on our experience, visual artifacts such as story maps and impact maps are well-suited, as they cannot be easily generated by applying GenAI tools. When students are asked to create user stories, it should be specified that these user stories are to be refined through collaborative discussion.

\textit{3. Reflection}. The third step involves critically reflecting on the learning process to strengthen the transfer of theory and practice. During a group discussion in which all teams are brought together, students reflect on their experiences with the stakeholder simulation, including the limitations of GenAI, and discuss ethical considerations related to the use of GenAI. Suitable discussion topics include:

\begin{itemize}
    \item Discussion of the quality of the artifacts produced for requirements specification and the application of agile practices (\textit{e.g.,} when applying user stories, the team discussion about the requirement is important for developing a shared understanding, and not only the production of the written part of the user story).
    \item The use of AI Personas for Agile RE has some limitations because of the bias in training data of the LLMs, which may result in stakeholder representations that do not reflect real target groups, leading to requirements that do not align with the needs of actual users. Though this aspect should be considered when applying AI Personas, we are aware of the limitation of a persona's mismatching to target groups when using the personas concept in general.
    \item Limitations of GenAI, such as losing contextual information, responding in a way that prioritizes efficiency over human-like communication, and the occurrence of hallucinations.
    \item Biased training data for LLMs (e.g., over-representation on one gender or under-representation of specific groups such as neurodivergent persons) should be covered.
    \item Depending on the module and the country in which the teaching case is embedded, it may also be appropriate to include a discussion on data privacy and regulations such as the EU AI Act and GDPR.
\end{itemize}

The teaching tips outlined above refer to the example of embedding the teaching case presented in Section~\ref{sec:EmbeddingTeachingCase}. The modular structure of the teaching case allows it to be easily adapted to different contexts, such as another module, a different case study with a different target group, or the use of alternative agile practices.

\section{Learnings and Experiences of the Practical Application}
\label{sec:PracticalImplications}

We have already applied the teaching case in various on-campus and online study programs and experimented with different configurations. This paper reflects our observations and experiences gathered during its implementations in different scenarios. On the one hand, we summarize our experiences related to the didactic design and the use of agile practices; on the other hand, we reflect on our experiences with the technical implementation of AI personas using generative AI tools.

\subsection{Learnings on Didactic Design and Use of Agile Practices}

We have applied the teaching case in different modules and across various study programs. In this context, we gained experience with the use of different agile practices for element 2. Requirements specification (see Figure~\ref{fig_TeachingCase}).

\textit{User Stories~\cite{Cohn2004}}. After completing the interviews, students used GenAI to generate user stories. In this step, additional advice from the lecturer was necessary to encourage students to discuss and refine the generated user stories within their groups. In several groups, it became apparent that the need for refinement was not initially recognized, as the GenAI-generated outputs sounded plausible and well-formulated. However, students did not realize that the user stories were too superficial for implementation in software development and lacked measurable acceptance criteria, e.g., documented as Defintion of Ready.

\textit{Story Maps~\cite{Patton2014}}. When students created story maps manually, the results were in line with the requirements of a story map. In contrast, when GenAI was used to generate story maps, two challenges became apparent. First, advanced prompting skills were required to obtain a visual story map template as output. Second, a story map is intended to tell a coherent story from the user’s perspective. With novice prompting skills, this proved difficult to achieve, as the GenAI-generated story maps tended to resemble collections of features grouped by category rather than user-centered narratives.

\textit{Impact Maps~\cite{NeuriConsultingLLP2026}}. When creating the impact map, some students experienced difficulties distinguishing between the concept of stakeholders used in the exercise and the concept of actors as defined by the agile practice of impact mapping. Contrary to the assumptions of some students, there is no one-to-one relationship between these concepts. In addition, the results produced by the student groups indicated challenges in identifying an appropriate level of granularity for the deliverables. Some groups listed non-functional requirements such as accessibility and user experience, while others focused on concrete features, such as user profiles or dashboards. When the exercise was conducted in September 2025, the LLMs in use were already capable of generating impact maps that met the requirements of the exercise based on the interview data, even with basic prompting skills. In this context, we recommend creating this artifact in an in-class setting, as it can otherwise be generated using GenAI without sufficient individual effort when assigned as homework.

Overall, it is advisable to revisit and critically reflect on the experiences students gain during the requirements specification phase in element 3. Reflection (see Figure~\ref{fig_TeachingCase}). Otherwise, there is a risk that students may apply agile practices incorrectly and produce low-quality results.

Additional feedback from students indicated that they initially struggled to understand their role within the task. We therefore recommend explicitly emphasizing, when introducing the assignment, that students assume the role of a requirements engineer who conducts interviews with stakeholders and is responsible for translating the elicited requirements into a specification.

\subsection{Learnings on Technical Implementation}

The LLMs provided by different GenAI providers are evolving rapidly. As a result, there are multiple ways to make stakeholder simulations using AI Personas available to students. In March 2025, we initially started with a customized-GPT hosted through OpenAI. Students were able to access the simulation via a direct link to the OpenAI platform. However, this approach led to several limitations related to licensing conditions. Students using a basic OpenAI account were only able to send a limited number of prompts to the chatbot (approximately five) and then had to wait several hours before continuing the interaction. This restriction significantly impaired the learning experience. In addition, some students expressed concerns regarding data protection and indicated a preference for using local GenAI models or alternative providers.
As a result, we transitioned to providing the stakeholder simulation through a meta-prompt. This approach allows students to decide independently which GenAI tool they wish to use to run the simulation. So far, we have had positive experiences with this approach.

To adapt the meta-prompt to newer LLMs, we optimized it in September 2025 using the Prompt Optimizer provided by OpenAI. This optimization noticeably changed the behavior of the AI Personas. Communication became strongly efficiency-oriented and less interactive. The AI Personas provided the requested information in a highly targeted manner and asked very few follow-up questions. Consequently, the AI Personas were perceived as less human-like, and the students finished their interviews quickly.

At present, we are experimenting with different versions of the meta-prompt to make the behavior of the AI Personas more human-like and to assign them distinct personalities. 

\section{Conclusion \& Future Work}
\label{sec:Conclusion}
This paper presents our learnings and experiences from a teaching case that integrates stakeholder simulation with AI Personas into Agile Requirements Engineering education in higher education. By combining interviews with stakeholders that are modeled as AI Personas with agile practices for requirements discovery and specification, our approach enables students to gain hands-on experience while critically reflecting on the technical and ethical limitations of generative AI. Our experiences from multiple course implementations indicate that AI Personas can serve as a valuable pedagogical instrument when used in a structured and reflective manner.

Future work will focus on further refining the meta-prompts used to control the behavior and personality of the AI Personas in order to support more natural and interactive stakeholder simulations. In this context, we are also evaluating the use of agentic AI. We are also experimenting with different forms of interaction, such as interviews via ChatGPT's voice interface. In addition, we plan to systematically evaluate learning outcomes across different course formats and student cohorts, as well as to explore extensions of the teaching case to other software engineering and project-based learning contexts.

\section*{Acknowledgements \& Final Remark}
The teaching material can be provided upon request. Finally, we note that text-parts of this paper were refined with the support of the GenAI tools ChatGPT (GPT-5.2) and DeepL.

\bibliographystyle{splncs04}
\bibliography{references}

@String{Computer = "{IEEE} Computer" }

@String{Springer = "Springer-Verlag" }

@inproceedings{Neumann2023,
   author = {Michael Neumann and Maria Rauschenberger and Eva-Maria Schön},
   note="doi: 10.1109/SEENG59157.2023.00010",
   booktitle = {5th International Workshop on Software Engineering Education for the Next Generation (SEENG)},
   pages = {29-32},
   title = {"We Need To Talk About ChatGPT": The Future of AI and Higher Education},
   year = {2023}
}

@misc{iso9241,
	author       = {{International Organization for Standardization}},
  title        = {ISO 9241-210:2019(en) Ergonomics of human-system interaction — Part 210: Human-centred design for interactive systems},
  year         = {2019},
  organization = {ISO},
  type         = {Standard},
  number       = {ISO 9241-210:2019(E)},
  address      = {Geneva, CH}
}

@INPROCEEDINGS{Bischof.2025,
  author={Bischof, L. and Schön, E.-M. and Rauschenberger, M. and Neumann, M.},
  booktitle={Proceedings of the 21st International Conference on Web Information Systems and Technologies}, 
  title={“We Need to Analyze Students GenAI Use”: Towards an AI Adoption Framework for Higher Education}, 
  year={2025},
  volume={},
  number={},
  pages={429-438},
  doi={10.5220/0013819000003985}}

@misc{OpenAI.2022,
 author = {OpenAI},
 year = {2022},
 title = {Chatgpt: Optimizing language models for dialogue.},
 url = {https://openai.com/blog/chatgpt/}
}

@book{Karimova.2025,
 author = {Karimova, Gulnara Z.},
 year = {2025},
 title = {Humanizing AI with Personality},
 address = {Cham},
 publisher = {{Springer Nature Switzerland}},
 isbn = {978-3-031-82326-8},
 doi = {10.1007/978-3-031-82327-5}
}

@article{Kollmorgen.2025,
title = {State of the art of Agile User Experience: Challenges and solution approaches},
journal = {Computer Standards {\&}  Interfaces},
volume = {96},
pages = {104083},
year = {2026},
issn = {0920-5489},
doi = {10.1016/j.csi.2025.104083}            ,
author = {Jessica Kollmorgen and Jenny Pilz and Tiago Silva {da Silva} and Jörg Thomaschewski and Eva-Maria Schön}}

@article{Sporsem.2025,
title = {User stories as boundary objects in agile requirements engineering: A theoretical literature review},
journal = {Journal of Systems and Software},
pages = {112693},
year = {2025},
issn = {0164-1212},
doi = {10.1016/j.jss.2025.112693},
author = {Tor Sporsem and Torgeir Dingsøyr and Klaas-Jan Stol}
}

@article{Schon2017f,
   author = {Eva-Maria Schön and Jörg Thomaschewski and María José Escalona},
   doi = {10.1016/j.csi.2016.08.011},
   issn = {09205489},
   journal = {Computer Standards \& Interfaces},
   month = {1},
   pages = {79-91},
   publisher = {Elsevier},
   title = {Agile Requirements Engineering: A systematic literature review},
   volume = {49},
   year = {2017}
}

@article{Fernandez2017,
   author = {D. Méndez Fernández and S. Wagner and M. Kalinowski and M. Felderer and P. Mafra and A. Vetrò and T. Conte and M. T. Christiansson and D. Greer and C. Lassenius and T. Männistö and M. Nayabi and M. Oivo and B. Penzenstadler and D. Pfahl and R. Prikladnicki and G. Ruhe and A. Schekelmann and S. Sen and R. Spinola and A. Tuzcu and J. L. de la Vara and R. Wieringa},
   doi = {10.1007/s10664-016-9451-7},
   issn = {15737616},
   issue = {5},
   journal = {Empirical Software Engineering},
   month = {10},
   pages = {2298-2338},
   publisher = {Springer New York LLC},
   title = {Naming the pain in requirements engineering: Contemporary problems, causes, and effects in practice},
   volume = {22},
   year = {2017}
}

@book{Cohn2004,
   author = {Mike Cohn},
   isbn = {978-0321205681},
   title = {User Stories Applied: For Agile Software Development},
   url = {https://books.google.com/books?hl=de&lr=&id=SvIwuX4SVigC&pgis=1},
   year = {2004}
}

@book{Anderson2001,
   author = {L.W. Anderson and D.R. Krathwohl},
   city = {New York},
   publisher = {Longman},
   title = {A Taxonomy for Learning, Teaching and Assessing: A Revision of Bloom’s Taxonomy of Educational Objectives: Complete Edition},
   year = {2001}
}

@book{Patton2014,
   author = {Jeff Patton},
   edition = {First edit},
   editor = {Mary Treseler and Amy Jollymore},
   isbn = {9781491904909},
   pages = {276},
   publisher = {O´Reilly},
   title = {User Story Mapping: Discover the Whole Story, Build the Right Product},
   year = {2014}
}

@misc{NeuriConsultingLLP2026,
   author = {Neuri Consulting LLP},
   title = {Impact Mapping},
   url = {https://www.impactmapping.org/},
   year = {2026}
}

@inproceedings{Pehar2025,
   author = {Franjo Pehar},
   doi = {10.1109/MIPRO65660.2025.11131969},
   isbn = {9798331535971},
   booktitle = {2025 MIPRO 48th ICT and Electronics Convention, MIPRO 2025 - Proceedings},
   pages = {1500-1509},
   publisher = {Institute of Electrical and Electronics Engineers Inc.},
   title = {AI as Synthetic Users in Human-Centred Design: A Systematic Review},
   year = {2025}
}

\end{document}